\documentclass[twoside]{article}
\usepackage{fleqn,espcrc2}
\usepackage{graphicx}

\newcommand{\beq}{\begin{equation}}
\newcommand{\eeq}{\end{equation}}

\def\csw{c_{\rm sw}}
\newcommand{\ksea}{\mbox{$\kappa^{\rm sea}$}}

\newcommand{\gtaeq}{\raisebox{-.6ex}{$\stackrel{\textstyle{>}}{\sim}$}}
\newcommand{\lesssim}{\raisebox{-.6ex}{$\stackrel{\textstyle{<}}{\sim}$}}

\title{Effects of non-perturbatively improved dynamical fermions in
UKQCD simulations}

\author{Alan C. Irving\address{Theoretical Physics Division,
         Department of Mathematical Sciences,
         University of Liverpool,
         PO Box 147, Liverpool L69 3BX, UK.},
	{\em UKQCD Collaboration}\thanks{supported by PPARC grants
	PPA/G/S/1998/00777 and GR/L22744} 
}

\begin{document}

\begin{abstract}
We present results for QCD with 2 degenerate flavours of quark
using a non-perturbatively improved action on a 
lattice volume of $16^3\times
32$ where the bare gauge coupling and bare dynamical quark mass have
been chosen to maintain a fixed physical lattice spacing and volume
(1.71 fm). By comparing measurements from these matched ensembles,
including quenched ones, we find
evidence of dynamical quark effects 
on the short distance static potential, the scalar glueball mass and
the topological susceptibility. There is little evidence of effects 
on the light hadron spectrum over the range of quark masses studied
($m_{\pi}/m_{\rho}\geq 0.60$). 
\end{abstract}

\maketitle

\section{SIMULATION STRATEGY}
\label{sec:strategy}
The first generation of full QCD simulations has confirmed that, for 
many purposes, the quenched approximation is indeed very good and is
hard to improve upon with finite computational resources. 
For recent reviews see~\cite{Burkhalter:1998wu,Mawhinney:2000fw}. 
Following an exploratory  series of simulations using
non-perturbatively improved Wilson fermions~\cite{Allton:1998gi}, the
UKQCD collaboration has now completed the first phase of a sequence of 
{\em matched simulations} in which some attempt has been made
to adjust the bare simulation parameters so that physically
relevant quantities 
remain approximately constant~\cite{Irving:1998yu}.
In particular, we have generated ensembles in which the Sommer
scale parameter $r_0$, expressed in lattice units, 
is matched as closely as is practical.

Table~\ref{tab:sim_params} contains a summary of the simulation
parameters and ensemble statistics. 
\begin{table}[htb]
\caption{Simulation parameters and statistics. 
Configurations are separated by 40 trajectories.}
\label{tab:sim_params}
\centering
\footnotesize
\begin{tabular}{ccclll} \hline \hline
$\beta$ &$\csw$ &$\ksea$ &Cfg. &$a$ [fm] &$m_\pi/m_\rho$ \\ \hline\hline
5.93    &1.82   &Quen.   & 624 &.1040(03) & - \\
5.29    &1.92   &.1340  &101 &.1018(10) &.830(07)\\
5.26    &1.95   &.1345  &101 &.1041(12) &.791(08)\\
5.20    &2.02   &.1350  &150 &.1031(09) &.688(10)\\
\hline \hline
\multicolumn{5}{l}{Lightest $\ksea$ simulations.\vspace{.15cm}}\\ \hline\hline
5.2     &2.02   &.13550 &208 &.0973(8) &.600(20)\\
5.2     &2.02   &.13565 &60 &.0941(8) &.567(50)\\
\hline \hline
\multicolumn{5}{l}{QCDSF\vspace{.15cm}}\\ \hline\hline
5.29     &1.92   &.13500 & &.0932(5) &.755(6)\\
5.29     &1.92   &.13550 & &.0889(13) &.694(9)\\
\hline \hline   
\end{tabular}
\normalsize
\vspace{-0.8cm}
\end{table}  
This strategy is motivated partly by pragmatism and partly by
theoretical considerations. With access to limited computational
resources, it was clear that the Collaboration would be unable 
to address immediately all
the limits required to recover a fully physical system: chiral,
continuum and infinite volume limits. 
We have therefore adopted a
{\em partially unquenched} approach in which, at finite lattice
spacing, the dynamical quark mass (two degenerate flavours) is treated
as an independently tunable parameter.
Rather than hold the bare coupling fixed,
we have chosen to hold $r_0/a$ fixed with the expectation that the
discretisation and finite size effects which are inevitably present
should be better under control. The choice of $r_0$ to set the scale 
removes the need to perform the chiral extrapolation
in valence quark mass which use of a non-gluonic observable
would entail. Of course an overall uncertainty in the phenomenological
value of $r_0$ remains ($\sim 2\%{}$), but the use of $r_0$ provides
a standard which is well-defined both theoretically and computationally. 
Here we use $r_0=0.49$ fm.

In the second phase, we are 
attempting to vary the lattice spacing while holding 
the ratio of pseudoscalar to vector meson mass fixed. 
Using fully non-perturbatively improved fermions~\cite{Jansen:1998mx}, we
expect that the discretisation errors are ${\cal O}(a^2)$.
The fact that the improvement coefficient function 
$c_{\rm SW}(\beta)$ is only well-determined for 
$\beta\geq 5.2$ limits the accessible lattice spacing from above.
The spacing is limited from below by the need
to have an acceptably large physical volume. 

\begin{figure}[htb]
\includegraphics[scale=0.425]{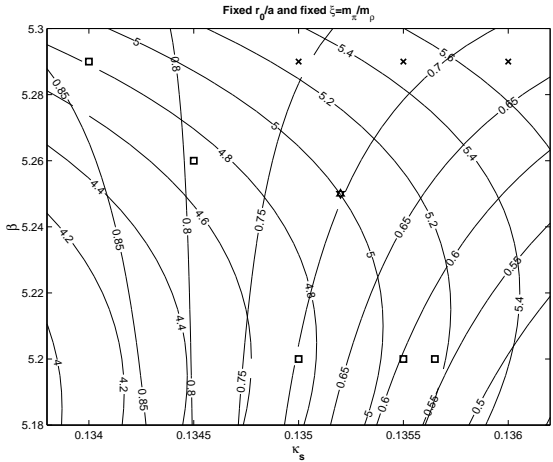}
\caption{Curves of constant $r_0/a$ (4, 4.2, \dots) 
and $\xi$ (0.5,0.55,\dots). 
The squares (star)
indicate completed (ongoing) UKQCD runs.
Crosses show ongoing and planned QCDSF runs.}
\label{fig:dublin}
\vspace{-0.5cm}
\end{figure}

We have begun a cooperative
programme of simulations with the QCDSF collaboration (see talks by
Pleiter and St\"uben~\cite{Pleiter_lat00,Stueben_lat00}). 
It is planned
that simulations on larger volumes and at complementary
values of the bare parameters will be obtained. 
Some initial simulations from QCDSF are also summarised in 
Table~\ref{tab:sim_params}. 
Simulation points are indicated in Figure~\ref{fig:dublin} which
has, superimposed on it, curves of constant $r_0/a$ and 
$\xi\equiv m_{\pi}/m_{\rho}$. 
These curves show a simple
polynomial interpolation which gives
a very approximate description of the data obtained so far and
provides guidance on the choice of future simulations points.

The lightest quark mass simulation which we have so far been able to
achieve with useful statistics was at
$(\beta,\kappa)=(5.20,0.1355)$ (see Table~\ref{tab:sim_params}).
We are currently generating
configurations which are expected to match the lattice spacing
of this ensemble and the value of $\xi$ for the ensemble
at $(5.20,0.1350)$. An attempt to simulate at $(5.20,0.13565)$ where
$\xi\approx 0.56$ (Table~\ref{tab:sim_params}) 
was halted since the performance of the
HMC algorithm was found to be unacceptably slow. 
For all ensembles studied, 
the measured integrated autocorrelation time for the plaquette (typically less
$\lesssim 20$ HMC trajectories) showed no increase with decreasing quark mass.
The same was true for measurements of the topological charge
and glueball correlators where the integrated autocorrelation time
was of the order of $25-30$.

\section{STATIC POTENTIAL}
\label{sec:pot}
\begin{figure}[htb]
\includegraphics[scale=0.35,angle=-90]{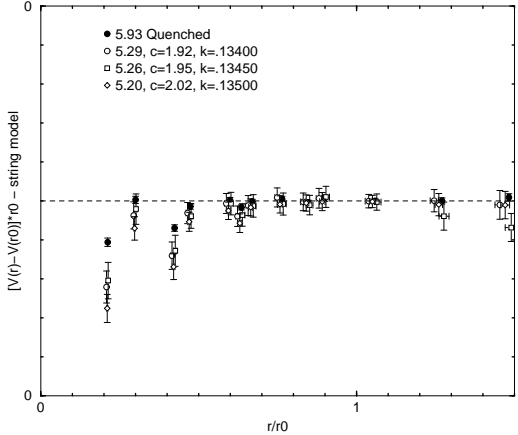}
\caption{Difference of the lattice potential from the
continuum bosonic string model.}
\label{fig:dpot}
\vspace{-0.5cm}
\end{figure}

The static potential was extracted using an optimised variational 
method~\cite{Michael:1985ne}. Within the range 
$0.35\,  \lesssim \, |{\bf r}|\, \lesssim \, 1.25$ fm, 
all matched data including quenched
are well described by a simple bosonic string model ($e=\pi/12$).
$$
[V(r)-V(r_0)]r_0=(1.65-e)\bigl({{r}\over{r_0}}-1\bigr) -
e\bigl({{r_0}\over{r}}-1\bigr)
$$
when expressed in units of $r_0$. Figure~\ref{fig:dpot} shows the
deviations from this model. 
For $r\, \lesssim \, 3a$ there is clear
evidence for (a) strong discretisation effects and (b) a systematic
depression of the potential with decreasing dynamical quark mass. 
Parametric fits show that the effect corresponds to
an increase in the effective strong coupling of $18\pm 10\%{}$
with respect to the quenched value. 
The simplest perturbation theory estimate of the effect of two
light flavours is also $18\%$. 

The values of $r_0/a$ used to 
obtain the effective lattice spacing shown in 
Table~\ref{tab:sim_params} were extracted from
parametric fits to the potential (including
off-axis measurements) but over a restricted
range of $|{\bf r}|$. In fact the values of $r_0/a$ are rather stable
with respect to the range used.

\section{MESONS and GLUEBALLS}
\label{sec:spect}
\begin{figure}[htb]
\includegraphics[scale=0.4]{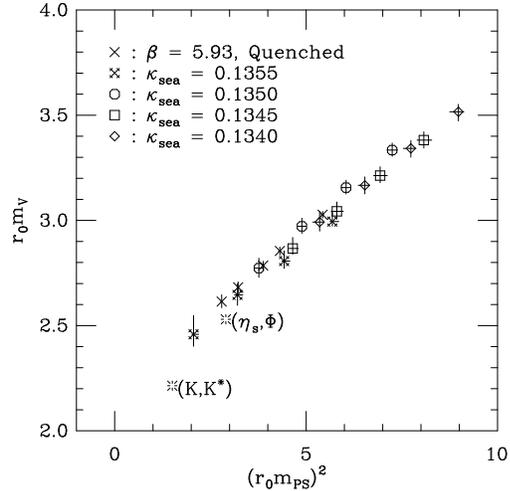}
\caption{Valence quark mass dependence of the vector mass for different
sea quark masses.}
\label{fig:mrhompi}
\end{figure}

Figure~\ref{fig:mrhompi} shows the valence quark mass dependence
($\sim m_{\rm PS}^2$) of the vector meson mass in comparison with
physical mesons having strange quark content. 
New matched quenched
data underline the conclusion that the lattice data are
not far from the physical data even at finite lattice spacing and
that there is little further improvement to be seen at the lightest
sea quark masses reached (corresponding to $m_\pi/m_\rho\approx 0.6$).

Results for baryons and for quark masses will be presented 
elsewhere~\cite{Allton:csw202}.

\begin{figure}[htb]
\includegraphics[scale=0.4]{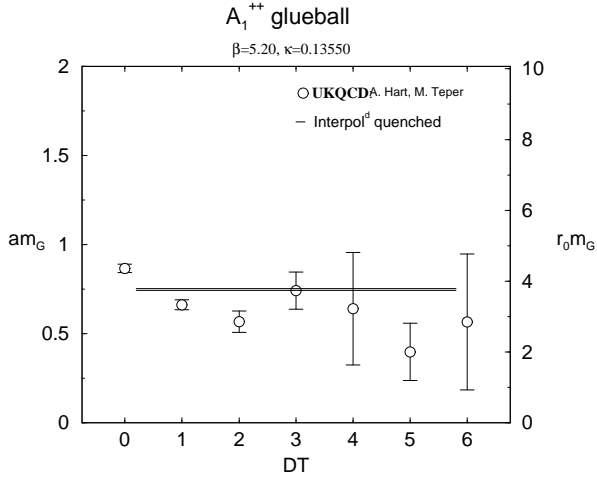}
\caption{Effective mass for the scalar glueball.}
\label{fig:gball}
\vspace{-0.5cm}
\end{figure}

In Figure~\ref{fig:gball} we show preliminary data for the 
mass of the scalar glueball ($A_1^{++}$ lattice state)
obtained using optimised variational techniques. 
The basis made use of 7 operators and 6 levels of fuzzing.
Another important ingredient was the use of non-zero momentum operators
$|{\bf p}|^2\leq 2a^2$) to obtain effective energies from which 
effective masses
could also be extracted. For these momenta, good plateaux were
again obtained. 

The figure shows the zero-momentum effective masses.
Data for the lightest quark ensemble 
($\ksea=0.1355$) are compared
with a \lq world average\rq{} of quenched scalar masses interpolated to 
the relevant lattice scale.  
This provides reasonably convincing evidence
of a decrease in the mass of order $10\%$ for the scalar glueball
at fixed lattice spacing. 

One possible source of (finite size) error could be mixing 
with a scalar \lq torelon\rq{} which has an additional decay
mode in the presence of fundamental charges.  
Direct measurements
of the energy and vacuum expectation values
of such states~\cite{Allton:csw202} show that the mixing
is expected to be small on these lattices.

Measurements of the tensor glueball have also been made
but larger statistical errors make the corresponding
comparison with quenched results less fruitful.

\section{TOPOLOGICAL SUSCEPTIBILITY}
\label{sec:topol}
\begin{figure}[htb]
\includegraphics[scale=0.4]{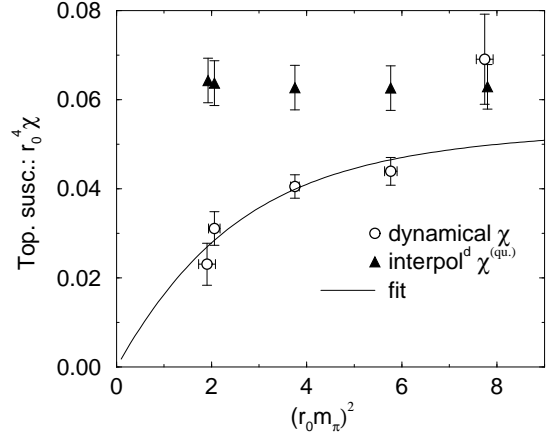}      
\caption{Scaled topological susceptibility versus effective quark mass.}
\label{fig:topol}  
\end{figure}

The matched ensembles have also been used to look for evidence of
suppression of instanton effects by light quark
modes~\cite{Hart:2000wr}.
In Figure~\ref{fig:topol} we show the suitably scaled
topological susceptibility obtained by standard cooling techniques
applied to the gauge configurations after every tenth trajectory.
Interestingly, the measured decorrelation rate of the topological charge was
little slower than for the plaquette.
The data show a clear suppression with dynamical quark mass. 
Since the leading chiral behaviour is expected to be
$$
\frac{r_0^2\chi}{m_\pi^2}
 =
(r_0 f_\pi)^2/8 + c_1 (r_0 m_\pi)^2\, ,
$$
one may use models to interpolate this data and hence 
extract an estimate of the pion decay constant
$f_\pi$. Note that, in contrast to methods based on matrix elements, 
no non-perturbative renormalisation factors
are required to obtain the value of $f_\pi$.
Using $r_0=.49$ fm to set the scale and ignoring finite-$a$
corrections, one estimates $f_\pi  =  148 \pm 7  ^{+25}_{-14}$ MeV
in comparison with $132$ MeV from experiment.

The use of matched data in this analysis is critical to its success in
that, otherwise, the underlying suppression can easily be masked by
the strong ($r_0^{-4}$) change in scale expected when changing
$\ksea$ at fixed $\beta$.

\section{SUMMARY and CONCLUSIONS}
\label{sec:concl}
The use of matched ensembles of dynamical fermion configurations 
has allowed a preliminary study of possible unquenching effects prior to
a full analysis of the continuum and chiral limits. Evidence is
presented for a lowering of the scalar glueball mass, charge screening
at short distances and of the suppression of instanton effects
at light quark masses. For the quark masses studied ($m_\pi/m_\rho
\, \gtaeq\, 0.6$) there is little evidence of dynamical quark effects in
the light hadron spectrum. Further simulations exploring the lattice
spacing dependence and quark mass dependence are underway.


\end{document}